\documentstyle[12pt]{article}

\begin{document}

\vskip 1cm

\begin{center}
{\bf THERMODYNAMICS OF THE TWO-DIMENSIONAL BLACK HOLE
IN THE TEITELBOIM-JACKIW THEORY} \\
\vskip 1cm
{\bf Jos\'e P. S. Lemos} \\
\vskip 0.3cm
{\scriptsize  Departamento de Astrof\'{\i}sica,
	      Observat\' orio Nacional-CNPq,} \\
{\scriptsize  Rua General Jos\'e Cristino 77,
	      20921 Rio de Janeiro, Brasil,} \\
{\scriptsize  \&} \\
{\scriptsize  Departamento de F\'{\i}sica,
	      Instituto Superior T\'ecnico,} \\
{\scriptsize  Av. Rovisco Pais 1, 1096 Lisboa, Portugal.} \\
\vskip 0.6cm
\end{center}

\bigskip

\begin{abstract}
\noindent
The two-dimensional theory of Teitelboim and Jackiw has constant and 
negative curvature. In spite of this, the theory admits a black 
hole solution with no singularities. In this work we study the 
thermodynamics of this black hole using York's formalism. 
\\

\end{abstract}

\newpage

\noindent
{\bf 1. Introduction}

\vskip 3mm

Thermal radiation from black holes via the Hawking process hints 
that gravity, quantum mechanics and thermodynamics are linked 
together. The analysis of quantum fields in a black hole background 
has first appeared in four dimensional (4D) general relativity. It 
was then extended to lower dimensions and other theories, following 
indications from string theory that these are important and useful to study. 

Two dimensions (2D) has been of particular interest after a black hole 
in string theory has appeared \cite{mandal,witten}. Hawking radiation and 
thermodynamics of this black hole has been analysed by several authors 
(e.g., 
\cite{callan,russo,frolov,fiola}). Another 2D theory which has been studied 
in some detail is the Teitelboim-Jackiw theory \cite{teitelboim,jackiw}. 
Although in this theory the curvature is constant and negative, 
it has a black hole solution 
\cite{christensenmann,achucarro,
cadonimignemi1,cadonimignemi2,lemossa1,lemossa2}.
The existence of a black hole implies a non-trivial causal strucuture 
which in turn generates interesting non-trivial thermodynamics. 
Hawking radiation of this black hole has been analysed in 
\cite{cadonimignemi2}, and thermodynamics of a black hole in a  
version of the theory with electromagnetic fields has been studied 
in \cite{kumar}. 

Here we study the black hole of the original Teitelboim-Jackiw theory 
using York's formalism \cite{york86,bradenetal}. In 2D this formalism 
has already been used in \cite{frolov} to study the 2D black hole 
in string theroy. The formalism uses the fact that for a system of 
fixed size and fixed temperature the canonical partition function 
$Z_c$ characterizes thermodynamic equilibrium in the canonical 
ensemble. The free energy $F$ and the partition function are linked 
through $-\beta F = \log Z_c$, where $\beta$ is the inverse temperature. 
On the other hand $Z_c$ can be 
represented by a path integral, through a relation with the Euclidean 
action $I_E$ given by $I_E=\beta F=-\log Z_c$. As a path integral, 
the partition function depends on quantities that are fixed in the 
functional integration such as the boundary data chosen from the fields 
of the system.

\vskip 1cm

\noindent
{\bf 2. The Lorentzian Black Hole Solution}

\vskip 3mm

In the Teitelboim-Jackiw 2D theory the action is 
\begin{equation}
I = \frac{1}{2\pi} \int d^2x\sqrt{-g} e^{\Phi}(R-2\Lambda) + I_B,
                         \label{eq:2.1}
\end{equation}
where $g$ is the determinant of the metric, $R$ is the curvature scalar, 
$\Lambda$ is the cosmological constant (sometimes written as 
$\Lambda=-2\lambda^2$), and $I_B$ is a boundary term to specify later. 
A 2D metric can always be written as
\begin{equation}
ds^2=-A({dx^0})^2 + \frac{({dx^1})^2}{P},
                         \label{eq:2.101}
\end{equation}
where $x^0$ and $x^1$ are the time and spatial coordinates, respectively,
and $A$ and $P$ are metric functions. 
The action (\ref{eq:2.1}) has got a black hole solution given in 
the unitary gauge by 
\begin{equation}
ds^2=-\sinh^2(\alpha x) dt^2 + dx^2,
                         \label{eq:2.2}
\end{equation}
\begin{equation}
e^\Phi = e^{\Phi_0} \cosh(\alpha x),
                         \label{eq:2.3}
\end{equation}
where the range of $x$ is $-\infty<x<+\infty$, $\Phi_0$ is a 
constant, and $\alpha^2 \equiv -\Lambda$. Transforming to the 
Schwarzschild gauge, through $r=\frac{\sqrt b}{\alpha}\cosh \alpha x$, 
with $b$ constant greater than zero, one obtains
\begin{equation}
ds^2=-(\alpha^2r^2-b)dt^2 + \frac{dr^2}{\alpha^2r^2-b},
                         \label{eq:2.4}
\end{equation}
\begin{equation}
e^\Phi=c\alpha r,
                         \label{eq:2.5}
\end{equation}
where $c$ is a constant. 

The maximal analytical extension of (\ref{eq:2.4}) (or (\ref{eq:2.2})) 
is represented in the Penrose diagram of figure 1. It is clear 
from that diagram and the metric given in equation 
(\ref{eq:2.4}) that the radius 
$r=\frac{\sqrt b}{\alpha}$ (or $x=0$) can represent a horizon. 
However, it can also be a coordinate trick. Indeed the curvature 
scalar of the solution is $R=-\alpha^2$ which is a constant. Therefore, 
spacetime has constant negative curvature and, in principle, is 
anti-de Sitter spacetime. Now, anti-de Sitter spacetime has, in the unitary 
gauge, a metric given by 
\begin{equation}
ds^2=-\cosh^2(\alpha x) dt^2 + dx^2,
                         \label{eq:2.6}
\end{equation}
and dilaton,
\begin{equation}
e^\Phi = e^{\Phi_0} \sinh(\alpha x).
                         \label{eq:2.7}
\end{equation}
To transform to the Schwarzschild gauge we put 
$\overline{r}=\frac{\sqrt{\overline{b}}}{\alpha}\sinh \alpha x$ and obtain 
\begin{equation}
ds^2 = - (\alpha^2\overline{r}^2 + \overline{b}) dt^2 
+\frac{d\overline{r}}{(\alpha^2\overline{r}^2 + \overline{b})},
                         \label{eq:2.8}
\end{equation}
\begin{equation}
e^\Phi = \overline{c} \alpha \overline{r}
                         \label{eq:2.9}
\end{equation}
where $\overline{c}$ is a constant, and $\overline{b}>0$ 
is also a constant which in this case can always be set to one, 
$\overline{b}=1$. The maximal analytical extension 
of (\ref{eq:2.8}) (or (\ref{eq:2.6})) is given by the usual 
anti-de Sitter extension \cite{hawkingellis}. 
It is then clear that $r$ and 
$\overline{r}$ are totally different coordinates. However, a set of 
transformations can indeed be found \cite{cadonimignemi2} as one might 
expect, since spacetime in both coordinates has constant negative 
curvature. Thus, in what sense can (\ref{eq:2.4})-(\ref{eq:2.5}) 
be interpreted as a black hole? Or, in other words, in what sense are 
(\ref{eq:2.4})-(\ref{eq:2.5}) and (\ref{eq:2.8})-(\ref{eq:2.9}) 
different physical solutions?

The interpretation of (\ref{eq:2.4})-(\ref{eq:2.5}) as a black hole comes 
from theories in 3D and 4D. It was shown in \cite{achucarro} that 
action (\ref{eq:2.1}) comes from dimensional reduction of 3D general 
relativity. 3D general relativity admits a static black hole solution 
with circular symmetry \cite{btz}. Solution (\ref{eq:2.4})-(\ref{eq:2.5}) 
gives the corresponding 2D black hole. In the 3D theory $e^{\Phi}$ is 
the circumference radius. On the other hand, it was also 
shown in \cite{cadonimignemi1} that (\ref{eq:2.1}) comes from 
dimensional reduction of a low energy 4D action of heterotic string theory, 
but now $e^{\Phi}$ represents instead the string coupling. This 4D 
action admits near-extremal magnetic black holes which in turn generate 
the ansatz for the dimensional reduction process. In both 3D and 4D theories 
it does not make sense in physical terms to have a negative $e^{\Phi}$. 
Thus, when (\ref{eq:2.1}) is used to model 3D and 4D black holes 
(as, for instance, s-wave scattering models in quantum evaporation 
of  black holes), one has to cut the 2D spacetime at $r=0$.
In both cases, it is the dilaton the field which sets this 
boundary condition. Therefore, solutions with the same local 
metric properties as in (\ref{eq:2.4})-(\ref{eq:2.5}) and 
(\ref{eq:2.8})-(\ref{eq:2.9}) are in fact topologically different. 
There is an extremal black hole solution given when the parameter 
$b=0$, see figure 3.

There is also the possibility of interpreting the solution 
(\ref{eq:2.4})-(\ref{eq:2.5}) as a black hole without having to resort 
to higher dimensions.  The idea in \cite{lemossa1} is that the line 
$-\infty<x<\infty$ (defined in  (\ref{eq:2.2})-(\ref{eq:2.3})) 
corresponds to the segment $\frac{\sqrt b}{\alpha}<r<\infty$. Each pair 
of space inverted points $(-x,x)$ degenerates into one $r$. A slice at 
constant (Penrose) time in the diagram of figure 1 is shown in figure 
3.  There is a horizon at $r=\frac{\sqrt b}{\alpha}$, i.e., $x=0$. 
Observers at each end of the line $x\rightarrow\pm\infty$ can only 
communicate if they enter through $x=0$. The $x=0$ segment is a null 
line, and test particles in timelike geodesics in one of the ends 
of the world ($x\rightarrow\pm\infty$) will cross this horizon in a 
finite time.  There is a problem in this interpretation. As figure 4 
indicates, there is a cusp (i.e., a singularity) at the junction 
$x=0$.  Observers (or particles) when entering a new world have to 
decide which end (positive or negative $x$) they will join.

Another 2D interpretation can be given to (\ref{eq:2.4})-(\ref{eq:2.5}).
One can notice that metric (\ref{eq:2.4}) represents a 
portion of the 2D anti-de Sitter spacetime in accelerated coordinates. 
Indeed, a stationary observer with $r=$constant in spacetime 
given by (\ref{eq:2.4}) has four acceleration $a^\mu$ with magnitude 
$a=\sqrt{a^\mu a_\mu}$ given by 
\begin{equation}
a = \frac{\alpha^2 r}{\sqrt{\alpha^2r^2-b}},
                         \label{eq:2.10}
\end{equation}
with $b>0$. 
The radius $r=\frac{\sqrt b}{\alpha}$, where the acceleration is infinite, 
corresponds to the trajectory of a light ray. Thus, observers 
held at $r=$constant see this light ray as a horizon, they will never 
see events beyond this ray. They are accelerated observers and 
can see only a portion of anti-de Sitter spacetime. 
In this sense, region II in figure 1, can be considered 
a black hole for region I accelerated observers. 
Note that for anti-de Sitter, $\overline{r}$=constant 
trajectories are straight vertical 
lines in the corresponding Penrose diagram \cite{hawkingellis}. 
In these coordinates the 
acceleration is $a = \frac{\alpha^2 r}{\sqrt{\alpha^2r^2+\overline{b}}}$, 
$\overline{b}>0$. 
There is no infinite acceleration for such observers.  
The situation is analogous to the relation that Rindler and Minkowski 
2D spacetimes bear with each other. 
However, here, there is an extra field, the dilaton.

Thus, equations (\ref{eq:2.4})-(\ref{eq:2.5}) represent a black hole 
in several different physical interpretations. In view of this 
it is interesting to show that this black hole solution has non-trivial 
thermodynamics. We use here 
the formalism developped by York \cite{york86,bradenetal} to understand 
the thermal behavior of the black hole, 
(for other types of formalism see  \cite{cadonimignemi2,kumar}). 

The mass of the black hole of equation (\ref{eq:2.4}) 
can be calculated by the standard procedures 
\cite{lemossa2} and is given by,
\begin{equation}
M=\frac{\alpha c}{2}b. 
                         \label{eq:2.11}
\end{equation}

\vskip 1cm

\noindent
{\bf 3. The Euclidean Black Hole and its Reduced Action}

\vskip 3mm

We now follow \cite{bradenetal,frolov} to find the 
reduced action of the system. 
We assume that there is a black hole inside a cavity with boundary $B$.
Now, the Euclideanized form of the metric (\ref{eq:2.101}) 
can be written as ($\eta=ix^0, \rho=x^1$),
\begin{equation}
ds_E^2 = A d\eta^2 + \frac{d\rho^2}{P}.
                         \label{eq:3.1}
\end{equation}
Here $\eta$ is a periodic coordinate running from $0$ to $2\pi$ and 
$\rho$ runs from $0$ at the horizon to $\rho_B$ at the boundary. The 
values of the metric function $A$ and dilaton $\Phi$ at the boundary are 
denoted by $A_B$ and $\Phi_B$. The inverse temperature $\beta$ at the boundary 
is related to the proper length of the boundary circle $S^1$ through the 
relation, 
\begin{equation}
\beta = \int_0^{2\pi} \sqrt A_B  d\eta=   2\pi \sqrt A_B.
                         \label{eq:3.2}
\end{equation}
The regularity conditions of the metric and dilaton fields at 
the horizon imply, 
\begin{equation}
\left.\sqrt{P} (\sqrt{A})'\right\rbrack_{\rho=0} =1
                         \label{eq:3.3}
\end{equation}
and
\begin{equation}
\left. P {\Phi'}^2 \right\rbrack_{\rho=0} = 0, 
                         \label{eq:3.4}
\end{equation}
where $'\equiv \frac{\partial}{\partial \rho}$.

The Euclidean action can be obtained from 
(\ref{eq:2.1}), 
\begin{equation}
I_{\rm E}= -\frac12 \int_V d^2x\sqrt{g} e^\Phi(R+\alpha^2) 
-\int_{\partial V} d\rho \sqrt h e^\Phi(K-K^0),
                         \label{eq:3.5}
\end{equation}
where the surface term is required to make the variational 
procedure self-consistent, which is important in analysing the 
thermodynamics, $h$ is the induced metric on the boundary, 
$K$ is the extrinsic curvature and $K^0$ is a 
term necessary to choose the background (the zero point energy). 
As before, $\alpha^2=2\lambda^2=-\Lambda$.
The equations of motion derived from (\ref{eq:3.5}) are, 
\begin{equation}
e^\Phi T_{ab} \equiv \frac12 D_a\Phi D_b\Phi + \frac12 D_a D_b \Phi 
- \frac12 g_{ab} D_c D^c\Phi + \frac12 g_{ab} D_c\Phi
D^c\Phi- \frac12 g_{ab}\alpha^2 =0.
                         \label{eq:3.6}
\end{equation}
Then the $T_{00}$ constraint, $T_{00}=0$, gives, 
\begin{equation}
\left\lbrack( P{\Phi'}^2-\alpha^2)e^{2\Phi}\right\rbrack'=0,
                         \label{eq:3.7}
\end{equation}
whose solution is 
\begin{equation}
P{\Phi'}^2-\alpha^2 = -\alpha^2 b e^{-2\Phi},
                         \label{eq:3.8}
\end{equation}
where  
we have chosen the constant of integration as $-\alpha^2 b$ 
appropriately. 
Now, using, 
\begin{equation}
\sqrt g R = - (\frac{\sqrt P A'}{\sqrt A})'\quad, \quad 
\sqrt g = \sqrt{\frac AP} \quad, \quad \sqrt h = \sqrt A 
\quad , \quad K=\frac12 \frac{\sqrt P A'}{A},
                         \label{eq:3.9}
\end{equation}
we can transform (\ref{eq:3.5}) into the following:
\begin{equation}
I_{\rm E} = -\frac12 \int d\eta d\rho e^\Phi 
(\frac{\sqrt P A'}{\sqrt A} \Phi' + \sqrt{A}{P} \alpha^2) 
-\frac12 \int d\eta e^\Phi 
\left.\sqrt\frac{P}{A}A'\right\rbrack_{\rho=0} + I_0,
                         \label{eq:3.10}
\end{equation}
where $I_0\equiv \int_{\partial V} d\rho \sqrt h e^\Phi K^0$
is an important term for choosing the background. 
Then, integrating (\ref{eq:3.10}) and using the constraints and boundary 
conditions we find, 
\begin{equation}
I(h^{-1}) = -(G^{-1})\beta e^{\Phi_B} \alpha \sqrt{1-e^{2(\Phi_H - \Phi_B)}}
-(h^{-1})2\pi e^{\Phi_H} + (G^{-1})\beta e^{\Phi_B} \alpha,  
                         \label{eq:3.11}
\end{equation}
where $\Phi_H$ is the 
value of $\Phi$ at the horizon and $I_0 \equiv \beta e^{\Phi_B} \alpha$ 
was chosen appropriately. 
In (\ref{eq:3.11}) we have put back Newton's constant $G$ and Planck's 
constant $h$ (still puting Boltzmann's constant and the velocity of the 
light equal to one). Note that in 2D we use the following 
units for the constants: $[G]=LM^{-1}T^{-1}$ and $[h]=MT^{-1}$. 
As in 4D \cite{bradenetal}, one sees that a quantum term has appeared in 
the action, namely the term $2\pi e^{\Phi}$, which is associated with 
the entropy of the system. 
Equation (\ref{eq:3.11}) is thus the reduced action 
$I=I(\beta,\Phi_B;\Phi_H)$ which yields the important 
thermodynamic quantities.

\vskip 1cm

\noindent
{\bf 4. Temperature and the Canonical Boundary Conditions} 

\vskip 3mm

To find the temperature we have to obtain the stationary point of the 
reduced action, by differentiating  $I(\beta,\Phi_B;\Phi_H)$ 
with respect to $\Phi_H$. Setting the resulting equation to zero, 
i. e., $\frac{\partial I}{\partial \Phi_H} = 0$, we find,
\begin{equation}
\beta = \frac{2\pi}{\alpha^2} \sqrt{W_B} e^{-(\Phi_H - \Phi_B)},
                         \label{eq:4.1}
\end{equation}
where,
\begin{equation}
W_B = \alpha^2 (1-e^{2(\Phi_H-\Phi_B)}).
                         \label{eq:4.2}
\end{equation} 
Equation (\ref{eq:4.1}) gives the inverse of the 
temperature($\beta=\frac{1}{T}$) of the 2D black hole. 

Now, a thermal equilibrium configuration in the canonical ensemble, has 
to yield $\Phi_H$ as a function of $\beta$. Indeed, inverting 
(\ref{eq:4.1}) gives 
\begin{equation}
\Phi_H = \Phi_B -\frac12 \ln(1 +\frac{\alpha^2\beta^2}{4\pi^2}) 
                         \label{eq:4.3}
\end{equation}
or in terms of the Schwarzschild gauge of equation (\ref{eq:2.4}) 
(where, $e^{\Phi_H} = \alpha r_H = \sqrt b = 
\sqrt{\frac{2 M}{\alpha c}}$) 
we find from (\ref{eq:4.3}), 
\begin{equation}
{\frac{2 M}{\alpha c}} = \alpha^2 r_H^2 = 
\frac{\alpha^2 r_B^2}{1+\frac{\alpha^2\beta^2}{4\pi^2}}.
                         \label{eq:4.4}
\end{equation}
Thus as $T\rightarrow0$ we have $M\rightarrow0$. As $T\rightarrow\infty$ 
we have a maximum mass $M_{\rm max} = \frac12 \alpha^3 c {r_B}^2$ for the 
BH in the thermal bath. That is, for a given $r_B$ 
the mass of the hole cannot be larger than the one which gives a 
horizon radius equal to $r_B$. 
There is nothing like the instanton solution of the Schwarzschild bath 
in 4D. 

In figure 5 we draw the graph, $r_H$ as a function of $r_B$. 
We see that, at equilibrium, for $T\rightarrow\infty$ one has $r_H=r_B$ for 
any $r_B$, 
while for $T\rightarrow0$ one has that $r_H$ is very small 
in relation to $r_B$.
This means that for very high temperatures, the boundary is located at 
the horizon, precisely. At low temperatures the boundary has to be 
far from the horizon radius. 

We now study some thermodynamic quantities in this 
canonical ensemble  formulation. We also analyse 
thermodynamic stability.

\vskip 1cm

\noindent
{\bf 5. Thermodynamical Quantities} 

\vskip 3mm

The entropy is defined through the equation
\begin{equation}
S_H = \beta \left(\frac{\partial I}{\partial \beta}\right)_{\Phi_B} - I.
                         \label{eq:5.1}
\end{equation}
Using (\ref{eq:3.11}) we find, 
\begin{equation}
S_H = 2\pi e^{\Phi_H},
                         \label{eq:5.2}
\end{equation}
which has the same functional expression as the one found in \cite{frolov}. 
In the Schwarzschild gauge it gives, 
\begin{equation}
S_H = 2\pi \sqrt{\frac{2 M}{\alpha c}}.
                         \label{eq:5.3}
\end{equation}
It is interesting to note that the functional dependence 
given in (\ref{eq:5.2}) is the same for all black holes having a simple 
2D Brans-Dicke action \cite{lemosprivate}.  Note also that the 
extreme case ($M=0$) has zero entropy. 

The thermodynamic energy $E$ is defined by
\begin{equation}
E\equiv \left.\frac{\partial I}{\partial \beta}\right)_{\Phi_B}.
                         \label{eq:5.4}
\end{equation}
Then from (\ref{eq:3.11}) we obtain 
\begin{equation}
E = \alpha e^{\Phi_B} - 
\alpha e^{\Phi_B} \sqrt{1-e^{2(\Phi_H-\Phi_B)}},
                         \label{eq:5.5}
\end{equation}
which, in the Schwarzschild gauge, can be put in the form
\begin{equation}
E = c \alpha^2 r_B \left(1 - \sqrt{1-\frac{{r_H}^2}{{r_B}^2}}\right).
                         \label{eq:5.6}
\end{equation}
We see here that the zero point was chosen so that when there is no 
mass ($r_H=0$) the thermal energy is zero. 
Since ${r_H}^2 = \frac{2 M}{\alpha^3 c}$ we can 
invert expression (\ref{eq:5.6}) to yield
\begin{equation}
\frac{1}{\alpha r_B} M = E - \frac{1}{\alpha^2 c} \frac{E^2}{2r_B},
                         \label{eq:5.7}
\end{equation}
which relates the ADM mass and the thermal energy. The ADM mass 
(the mass at infinity) is equal to the termal energy times 
the length (in intrinsic units) of the reservoir minus a self-energy 
thermal term. Expression (\ref{eq:5.7}) is the closest one can get 
to the Schwarzschild expression found in \cite{york86} for the 
Schwarzschild mass, i.e., $M= E - \frac12 \frac{E^2}{r_B}$. 

Now, we want to find the Euler relation for this thermodynamic system. 
From (\ref{eq:4.1}) we obtain the temperature $T=\frac1\beta$, 
\begin{equation}
T = \frac{\alpha}{2\pi} \frac{r_H}{r_B} 
\frac{1}{\sqrt{1-\frac{{r_H}^2}{{r_B}^2}}}.
                         \label{eq:5.8}
\end{equation}
We define a linear pressure by
\begin{equation}
p = - \frac{\partial E}{\partial r_B} = 
\left( 1 - \sqrt{1-\frac{{r_H}^2}{{r_B}^2}}\right).
                         \label{eq:5.9}
\end{equation}
Then, using (\ref{eq:5.3}), (\ref{eq:5.7}), (\ref{eq:5.8}) 
and (\ref{eq:5.9}) we obtain 
\begin{equation}
dE = TdS - pdr_B.     
                        \label{eq:5.10}
\end{equation}
After integration we obtain the Euler relation
\begin{equation}
E = TS - pr_B.
                         \label{eq:5.11}
\end{equation}
Upon scaling, $r_B\rightarrow lr_B$ and $r_H\rightarrow lr_H$ 
or ($S\rightarrow lS$) one has $E\rightarrow lE$. Thus, $E$ is 
homogeneous of degree 1 in $S$ and $r_B$. 

To analyse thermodynamic stability we first find the heat capacity.
For 2D black holes it is defined by
\begin{equation}
C_{r_B} \equiv T \left( \frac{\partial S}{\partial T} \right)_{r_B}
                         \label{eq:5.12}
\end{equation}
Using the expressions (\ref{eq:5.2}) for $S_H$ we find
\begin{equation}
C_{r_B} = 2\pi c \alpha \frac{r_H}{{r_B}^2} ({r_B}^2 - {r_H}^2).
                         \label{eq:5.13}
\end{equation}
Thus the heat capacity is positive always, since $r_B\ge r_H$. Therefore, 
one has thermal stability always. The root-mean-square energy fluctuations 
$\Delta E$ are given by
\begin{equation}
<(\Delta E)^2 > = C_{r_B} T^2 = \frac{\alpha^3 c}{2\pi} 
\frac{{r_H}^3}{{r_B}^2}.
                         \label{eq:5.14}
\end{equation}
When $r_B \rightarrow r_H$ we have $\Delta E$  finite  and given by
$\sqrt{<(\Delta E)^2>} =\sqrt{ \frac{\alpha^3 c}{2\pi} r_H }$. 
\vskip 1cm

\noindent
{\bf 6. Free Energies and the Ground State of the Canonical Ensemble} 
\vskip 3mm

The Helmholtz free energy function for black holes, $F_{BH}$, can be deduced 
from the action by the relation
\begin{equation}
I_{BH} = \beta F_{BH}.
                         \label{eq:6.1}
\end{equation}
This free energy applies to the equilibrium value of the 
mass (or $r_H$) given in (\ref{eq:4.4}). From (\ref{eq:3.11}) 
we have in the Schwarzschild 
gauge the following free energy for the black hole, 
\begin{equation}
F_{BH} = - \alpha e^{\Phi_B} \frac{r_B - \sqrt{{r_B}^2 - {r_H}^2}}
{\sqrt{{r_B}^2 - {r_H}^2}},
                         \label{eq:6.2}
\end{equation}
which is non-positive for all $r_B$. Then, the action at equilibrium 
is given by the equation, 
\begin{equation}
-I_{BH} =  \alpha e^{\Phi_B} \beta\frac{r_B - \sqrt{{r_B}^2 - {r_H}^2}}
{\sqrt{{r_B}^2 - {r_H}^2}}.
                         \label{eq:6.3}
\end{equation}
But from (\ref{eq:4.1}) the inverse temperature is given by
$
\beta = \frac{2 \pi}{\alpha} \frac{r_B}{r_H} 
{\sqrt{1-\frac{{r_H}^2}{{r_B}^2}}},
$
which can be inverted to yield the relation, 
$
\frac{{r_H}^2}{{r_B}^2} = 
\frac{1}{\frac{\alpha^2\beta^2}{4\pi^2}+1}$. 
Then (\ref{eq:6.3}) can be put in the form
\begin{equation}
-I_{BH}(r_H) = -\beta e^{\Phi_B} \alpha + 2\pi e^{\Phi_B} 
\sqrt{1+\frac{\alpha^2\beta^2}{4\pi^2}}.
                         \label{eq:6.4}
\end{equation}

We now find the free energy for hot anti-de Sitter space (HADS) in 2D.
The local energy density, $\rho_0$, 
of radiation can be found to be 
\begin{equation}
\rho (T) = \frac{\pi}{12} g {T_{\rm local}^2},
                         \label{eq:6.5}
\end{equation}
where $g$ is the number of massless spin sates 
and where $T_{\rm local}$ is the locally measured temperature. 
The energy-momentum tensor of a perfect fluid is 
\begin{equation}
T_{ab} = \rho u_a u_b + p (g_{ab} + u_a u_b). 
                         \label{eq:6.6}
\end{equation}
A perfect radiation fluid in 2D obeys the following equation of 
state
\begin{equation}
p=\rho.
                         \label{eq:6.7}
\end{equation}
Thus in 2D the energy-momentum tensor of radiation becomes, 
\begin{equation}
{T^a}_b = \rho ({\delta^a}_b -  2  {\delta^a}_0 {\delta^0}_b ).
                         \label{eq:6.8}
\end{equation}
Therefore, 
\begin{equation}
{T^0}_0 = -\rho = - \frac{\pi}{12} g {T_{\rm local}}^2. 
                         \label{eq:6.9}
\end{equation}
By the 
Tolman formula we have
\begin{equation}
T_{\rm local}  = \frac{T}{\sqrt{-g_{00}}},
                         \label{eq:6.10}
\end{equation}
where $T$ is the temperature measured at infinity. 
Thus, (\ref{eq:6.7}) yields
\begin{equation}
- {T^0}_0 = \rho =  \frac{\pi}{12} g \frac{{T}^2}{(-g_{00})}.
                         \label{eq:6.11}
\end{equation}
The Tolman 
energy for HADS can also be found, 
\begin{equation}
E_{\rm HADS} = \int \rho dV = \int \rho \sqrt{-g} dx.
                         \label{eq:6.12}
\end{equation}
where $V$ is the proper volume of the energy one wants to measure. 
Now, in the Schwarzschild gauge, anti-de Sitter spacetime has metric 
given by (\ref{eq:2.8}). Then, $\sqrt{-g}=1$. Thus
\begin{equation}
E_{\rm HADS} = \int \rho dx = {\int_{r_B}}^{r_B} 
\frac{f(T)}{(-g_{00})} dr = f(T) {\int_{r_B}}^{r_B} 
\frac{dr}{\alpha^2 r^2 +1} = f(T) \overline{V}.
                         \label{eq:6.13}
\end{equation}
Here, $\overline{V}$ is the optical volume of radius $r_B$, defined by, 
\begin{equation}
\overline{V} = \int_{-r_B}^{r_B}  \frac{dr}{\alpha^2 r^2 + 1}  = 
\frac{2}{\alpha} {\rm arctan} (\alpha r_B).
                         \label{eq:6.14}
\end{equation}
We see here that ADS spacetime behaves as an enclosure of finite volume. 
From (\ref{eq:6.14}) we have,
\begin{equation}
E_{\rm HADS} = f(T) \frac{2}{\alpha} {\rm arctan} (\alpha r_B) = 
\frac{\pi}{6\alpha} g {T}^2 {\rm arctan} (\alpha r_B).
                         \label{eq:6.15}
\end{equation}
For $\alpha r_B \rightarrow \infty$ one has, 
$E_{\rm HADS}(r_B\rightarrow \infty) = \frac{\pi^2}{12\alpha} g {T}^2$, 
which is the energy for the whole spacetime. The 
action for HADS can be taken from the expression, 
$I_{\rm HADS} =  \int E_{\rm HADS} d\beta$. Using (\ref{eq:6.15}) one 
obtains,
\begin{equation}
-I_{\rm HADS} = \frac{\pi}{6\alpha} g T {\rm arctan} (\alpha r_B). 
                         \label{eq:6.16}
\end{equation}

The ground state is the state of least free energy. 
Since $I=\beta F$, and $\beta 
\geq 0$, we can compare directly the reduced actions for HADS and 
the black hole. 
We find that HADS dominates whenever
\begin{equation}
I_{\rm HADS} \leq  I_{\rm BH},
                         \label{eq:6.17}
\end{equation}
Then using equations (\ref{eq:6.4}) 
and (\ref{eq:6.16}) one obtains, 
\begin{equation}
T\geq \alpha \frac{12c}{g} 
\frac{\alpha r_B}{{\rm arctan (\alpha r_B)}}
\sqrt{1-\frac{g}{12\pi c} \frac{{\rm arctan(\alpha r_B)}}{\alpha r_B}}.
                         \label{eq:6.18}
\end{equation}
Whenever the number of particle species is relatively large then HADS 
is favoured for sufficiently small $r_B$. Indeed, if $g>12\pi c$, then 
the quantity inside square brackets is negative up to some 
boundary radius given 
implicitly by $\frac{\alpha r_B}{\rm arctan(\alpha r_B)}= \frac{12\pi 
c}{g}$. This means that up to this radius HADS dominates and for larger 
$r_B$ HADS dominates if $T$ obeys (\ref{eq:6.18}) (see figure 5, line (a)). 
If $g<12\pi c$ then HADS is favoured only if $T$ obeys (\ref{eq:6.18}) (see 
figure 5, line(c)). 
The case $g=12\pi c$ says that for $r_B\rightarrow0$ HADS is 
dominant (see figure 5, line (b)). 
Note that when 
the boundary $r_B\rightarrow \infty$ one obtains that, for finite 
temperature, the black hole is the ground state. 

It is also interesting to find the density of states, $\nu(E)$.  
Following \cite{york86}, one finds 
\begin{equation}
\nu(E) = \delta(E-A) e^{2\pi \Phi_H}.
                         \label{eq:6.19}
\end{equation}
Thus the density of states is proportional to the entropy. 
\vskip 1cm

\noindent
{\bf 7. Conclusions} 

\vskip 3mm

The Teitelboim-Jackiw theory has, 
in absence of matter, constant curvature spacetime solutions. 
Therefore the black hole 
solution of the theory has no singularities. In the first studies 
exploring this theory  it was thought 
that such a black hole did not exist. However, solutions containing 
point particles and horizons were found \cite{brownhenneauxteitel} 
which also had some 
interesting thermodynamic properties. To establish the existence of 
the black hole in this theory one has to invoke topological arguments. 
This  solution is special in the sense that to  
have a black hole one needs to add features which are not contained 
in the metric, i.e., one has to add boundary conditions.

We have then showed that this black hole yields non-trivial thermodynamics 
in York's scheme. Through an analysis of the free energies of both 
the black hole solution and hot anti-de Sitter spacetime it was 
possible to infer that for small enough ambient temperature the 
black hole is the ground state.

\vfill\eject
\centerline{Figure Captions}
\vskip 1cm

Figure 1 - The Penrose diagram for  metric (\ref{eq:2.4}). 

Figure 2 - The Penrose diagram for the non-singular black hole.

Figure 3 - The Penrose diagram for the extreme black hole. 

Figure 4 - The line $-\infty<x<\infty$ has a junction at $x=0$ 
(or $r=r_H$). Observers on each side of the line can only 
communicate if they cross $x=0$. The time direction is vertical. 

Figure 5 - The horizon radius is plotted as a function of the 
radius of the boundary for a given temperature, see equation 
(\ref{eq:4.4}). For each temperature the line is straight. 
It is also shown which regions favour hot anti-de Sitter 
spacetime and which favour the existence of a black hole, 
see equation (\ref{eq:6.18}).
When $g>12\pi c$ (in the figure it was used $\frac{g}{12\pi c}=2$), 
HADS is favoured to the left of line (a), (this case is 
represented in this figure). 
When $g=12\pi c$, HADS is favoured to the left 
of line (b).
When $g<12\pi c$ (in the figure it was used $\frac{g}{12\pi c}=\frac12$), 
HADS is favoured to the left of line (c). See text for more details.

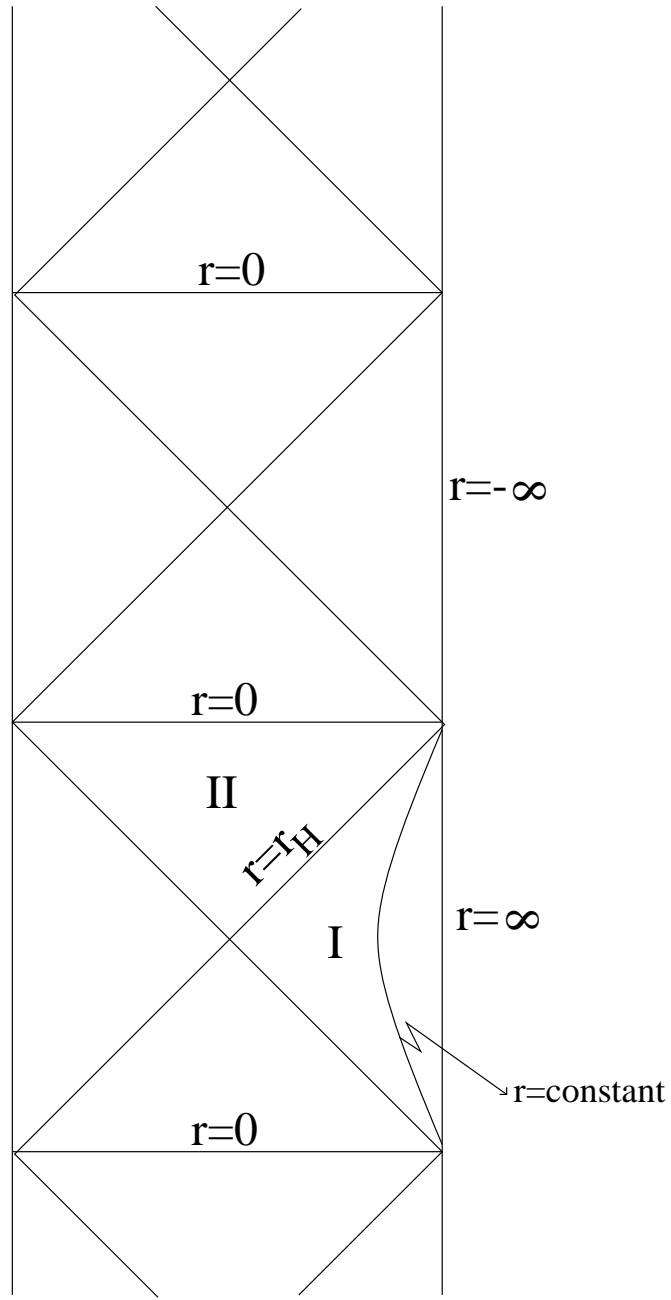
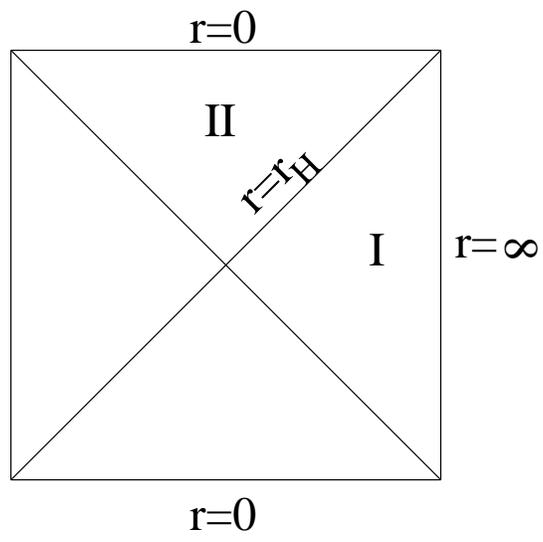
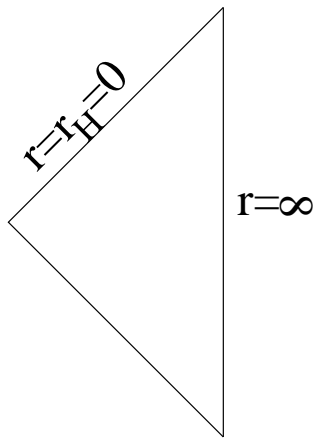
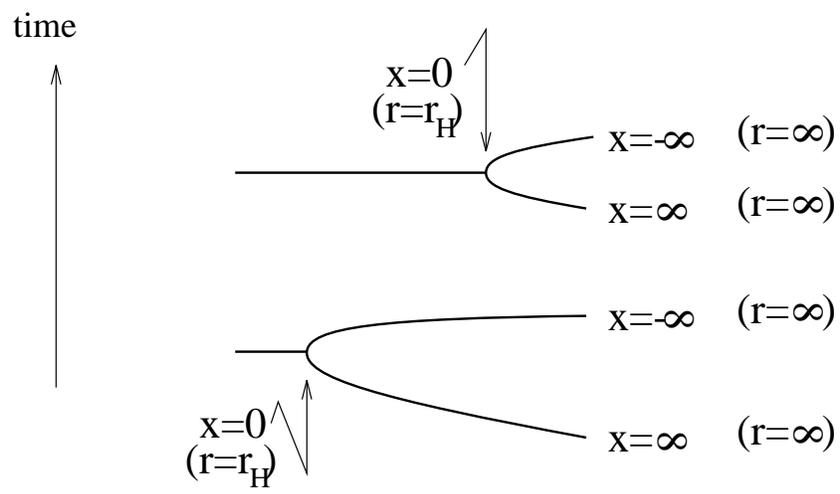
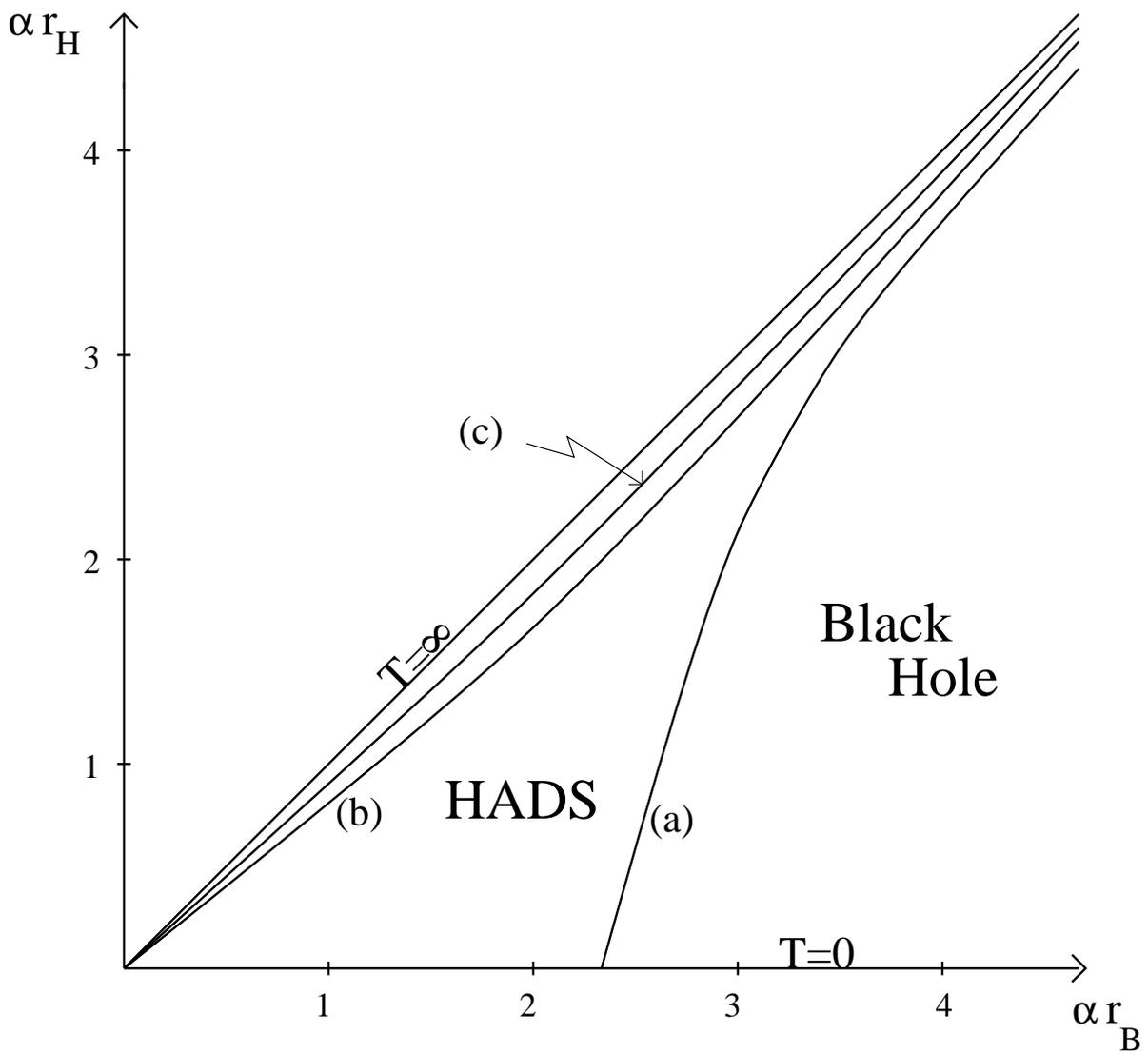
\end{document}